\documentclass{IEEEtran} 
\usepackage{times,amssymb,amsmath,graphicx,verbatim}

\title{Lattices for Physical-layer Secrecy: \\ A Computational Perspective}

\author{
Fabio Fernandes \& Sriram Vishwanath \\ Lab. of Informatics, Networks and Communications (LINC)\\
Dept. of Electrical and Computer Engineering \\University of Texas, Austin\\
\{fabio,sriram\}@austin.utexas.edu
}

\begin{document}

\maketitle

\begin{abstract}
In this paper, we use the hardness of quantization over general lattices as the basis of developing a physical layer secrecy system. Assuming that the channel state observed by the legitimate receiver and the eavesdropper are distinct, this asymmetry is used to develop a cryptosystem that resembles the McEliece cryptosystem, designed to be implemented at the physical layer. We ensure that the legitimate receiver observes a specific lattice over which decoding is known to be possible in polynomial-time. while the eavesdropper observes a lattice over which decoding will prove to have the complexity of lattice quantization over a general lattice\footnote{This work was supported in part by a Brazilian national fellowship}.

\end{abstract}

\section{Introduction}

Security is and will remain one of the primary design requirements in any communication system. Depending on the nature of security desired, there are multiple means by which secure communication is enabled \cite{Liang, Rhee}. An increasingly important means of securing communication is by exploiting asymmetries (between the legitimate pair and the wiretapper) at the physical layer. There are many schemes already in existence designed to enable physical layer security \cite{Physical,Liang}. In general, there are two settings in which physical layer security is conventionally studied: information-theoretic and computational.

\begin{figure}[ht]
\begin{center}
\includegraphics[width=6.7cm]{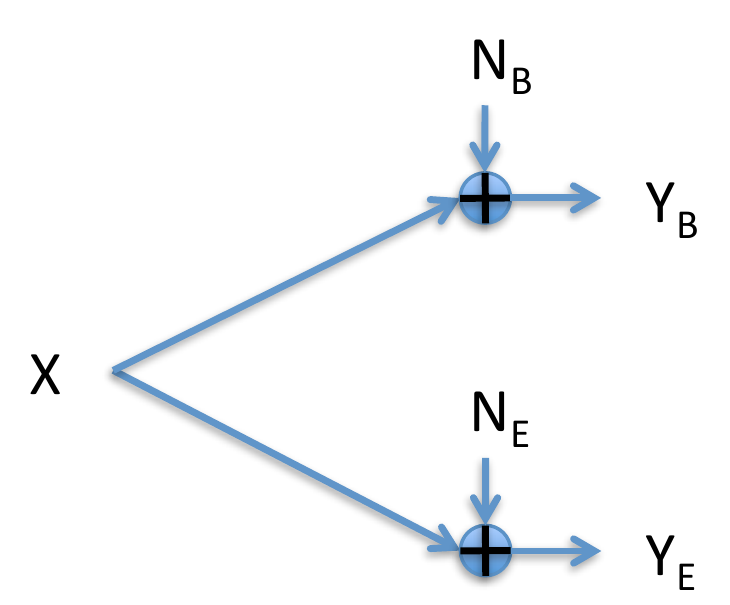}
\caption{The channel model: $X$ is the legitimate source while $Y_B$ is the signal observed by the legitimate receiver. $Y_E$ is observed by the eavesdropper}
\label{fig:model}
\end{center}
\end{figure}

From the information-theoretic perspective, the Wyner wiretap model is the arguably one of the best studied secure-communication models \cite{Wyner}. In this setting, the legitimate transmitter (Alice) wishes to communicate a message $W$ to a legitimate receiver (Bob) through a noisy channel. An eavesdropper (Eve) is present that can overhear the communication through another noisy medium. This wiretap model is depicted in Figure \ref{fig:model}. The notion of secrecy employed here is typically that of {\em perfect} secrecy, i.e., the eavesdropper is not assumed to be computationally bounded and we desire that there is absolutely no leakage of information. However, the results obtained using the Wyner wiretap model can be fairly pessimistic - if the wiretapper has a ``better" channel than the legitimate receiver, the capacity of this model is zero.

Given the stringent nature of the information-theoretic perfect secrecy, the notion of computational secrecy (which forms the basis of many cryptosystems) has received significant attention \cite{Rhee}.  Computational secrecy aims to create a computational asymmetry in the system, enabling the legitimate receiver to determine the message at low complexity while ensuring that it is exponentially complex for the eavesdropper to do so. Indeed, a large number of secure systems deployed today use a cryptosystem based on computational asymmetry. However, a majority of these cryptosystems operate at layers higher than the physical layer. Although mechanisms for exploiting the physical layer for computational secrecy have been studied (\cite{Physical} and references therein), a not-so-uncommon mindset is to use cryptosystems at higher layers and to assume that  perfect-secrecy is only relevant the physical layer.

In this paper, we utilize channel-asymmetry in enabling computational secrecy at the physical layer using lattices. Our approach is similar to that of a McEliece cryptosystem. A McEliece cryptosystem is based on the difficulty of syndrome decoding for a general linear codes, which is known to be NP hard \cite{McEliece}. However,  decoding of particular classes of codes (such as BCH or Goppa) is known to be possible with low complexity. Thus, a McEliece cryptosystem is designed so that the legitimate receiver observes a very specific code while the eavesdropper is faced with a general linear code. The McEliece cryptosystem, however, does not use channel asymmetry as its basis, and requires key exchanges making it is a higher (network or application)-layer cryptosystem. This this paper, we build a cryptosystem that does not require key exchanges and is designed to operate at the physical layer. We use lattices and the fact that lattice decoding of a general lattice is NP-hard \cite{Khot} to build this physical-layer cryptosystem. Indeed, the shortest vector problem (SVP) in lattices has already been used to develop cryptosystems \cite{Peikert}. However, these cryptosystems typically operate at higher layers, and it is our goal to bring this to the physical layer. In short, the main idea is to encode Alice's transmission such that Bob observes a lattice that is easy to decode while Eve observes a general lattice over which lattice quantization is exponentially hard.

Lattices for communication over additive Gaussian noise (AGN) channels have already been studied in detail \cite{Zamir}. Elements from the ensemble of Construction-A lattices \cite{Conway} have been shown to be optimal for the AGN channel \cite{Rimoldi,Erez}. However, general Construction-A lattices do not lend themselves to low complexity encoding and decoding algorithms. Thus, specific lattice structures are now being studied to enable lattice-based communication. One such effort is the design of low density lattice codes (LDLCs) \cite{LDLC}, while another is a specific structured design of block-triangular Construction-A lattices \cite{Ghiya}. In this paper, we use one of these specific lattices as the communication lattice to the legitimate receiver (Bob), and ``scrambling" it so a general lattice is observed at the eavesdropper.

The rest of this paper is structured as follows: the next section presents the system model. Section \ref{sec:prelim} summarizes some lattice preliminaries. In Section \ref{sec:main}, we present two mechanisms for lattice-based computational secrecy - the first based on simple channel inversion and the second on channel diagonalization and inverse water-filling. Section \ref{sec:block} describes a particular lattice construction based on a block-lower-triangular scheme as presented in \cite{Ghiya}, while  Section \ref{sec:conclude} concludes the paper.

\section{System Model}
\label{sec:model}
A quick note on notation. For a matrix $H$, $H^{-1}$ denotes its inverse and $H^t$ its transpose. $X^n$ denotes an $n \times 1$ vector. Oftentimes, the vector $X^n$ is abbreviated as $X$.

As mentioned in the introduction, the Wyner wiretap model is considered in this paper. There is one transmitter, Alice, that wishes to communicate with  Bob. Eve is a wiretapper that must be denied access to the information being transmitted. The channel is an intersymbol interference (ISI) channel that can be written in matrix form as

\[
Y_B = H X + N_B\]

where $X$ is an $n \times 1$ vector formed by the signal transmitted by Alice and $Y_B$ is a $n \times 1$ signal observed by the legitimate receiver. $H$ is an $n \times n$ matrix corresponding to the legitimate channel. The channel is assumed to be fairly general, with time-variation and/or ISI. Indeed, it is important that $H$ not be just a scaled version of the identity matrix\footnote{This paper utilizes the time-varying/ISI nature of the medium, and the arguments do not hold if it is a constant channel.}.

The channel over $n$ time-instances to the eavesdropper is given by
\[
Y_E = G X + N_E
\]

where $G$ is an $n \times n$ matrix that is independently generated and thus distinct from $H$. To make this notion of independence precise, a special case would be the case where the entries of both $H$ and $G$ are drawn i.i.d. over a suitable distribution.

\subsection{Assumptions}
The main assumption is that the transmitter and legitimate receiver know the channel transformation $H$. The fact that $H$ and $G$ are two different matrices is the main asymmetry being used in this design. $H$ can be estimated in a wireless time division duplex (TDD) setting by both the sender and receiver. In a frequency division duplex (FDD) setting, a feedback mechanism is needed to ensure that the channel estimate is available to the transmitter. A couple of points:
\begin{enumerate}
\item $G$ is {\em not} assumed to be known to either Alice or Bob. Even the presence of an eavesdropper need not be known to the legitimate pair. It suffices that $G$ and $H$ be ``sufficiently" different.
\item The channel state of the legitimate pair $H$ can be known to the eavesdropper, and even then the secrecy results obtained are meaningful.
\end{enumerate}

\subsection{Lattice Preliminaries}
\label{sec:prelim}

A lattice $\Lambda$ is a collection of vectors in ${\mathbb R}^n$, of the form:
\[
\Lambda = \{\lambda = Gx, x \in {\mathbb Z}^n\}
\]

where ${\mathbb Z}^n$ is the integer lattice, a collection of all integers vectors of length $n$ and $G$ is an $n \times n$ real-valued matrix.  Let $\Omega$ denotes the fundamental Voronoi region of the lattice $\Lambda$ and ${\mathcal V}$ denotes the volume of $\Omega$. There are two figures of merits for lattice: The volume to noise ratio (VNR)  and the normalized second moment (NSM), whose definitions can be found in \cite{Zamir}. The VNR measures a lattices suitability for communication over AGN channels, while the NSM the same for compression of Gaussian sources with a squared distortion measure. Overall, ``good" lattices with respect to both VNR and NSM are known to exist. These ``good" lattices, are also known to achieve the capacity of an additive Gaussian noise (AGN) channel are known to exist \cite{Erez}.

Next, we proceed to describing the physical-layer lattice based cryptosystem envisioned in this paper. The lattices used for communication may or may not be ``good" for source or channel coding. They will, however, be structured to ensure efficient decoding and the legitimate receiver while making decoding at the eavesdropper difficult.

\section{Lattice-based Cryptosystem}
\label{sec:main}

The design of a lattice based cryptosystem can be attained in a  fairly straightforward manner, along the same lines as the McEliece cryptosystem. 
\subsection{Channel Inversion}

The simplest construction is when the matrix $H$ is invertible. In this construction, if $\Lambda$ is a lattice that  enables low complexity encoding and decoding at the legitimate pair. Then, the following policy is used:

 \noindent{\bf Encoding}: map message $m$ to a lattice point $\lambda \in \Lambda$. This is possible due to the low-complexity nature of encoding associated with $\Lambda$.
\vspace{0.3mm}

\noindent{\bf Transmission:} Alice communicates 
\[
X = C H^{-1} \lambda
\]
where $C$ is a suitable normalization constant so that the power constraint at the transmitter is met.

\noindent{\bf Decoding:} Bob observes $Y_B^n = C \lambda + N_B^n$. Given the structure of $\Lambda$, decoding this to recover $\lambda$ and thus the message $m$ is possible with polynomial complexity.

 \noindent{\bf Eavesdropper:} The eavesdropper observes:
\[
Y_E^n = C G H^{-1} \lambda + N_E^n
\]
If  $G$ and $H$ are independently generated with i.i.d. entries over a continuous valued distribution, it is easy to verify that the  probability that $G H^{-1}$ is a  unitary matrix goes to zero as $n \rightarrow \infty$. This implies that the new lattice $G H^{-1} \lambda$ no longer has the desired low-complexity structure and thus cannot be decoded using polynomial-time algorithms. Note that, if $H$ is unknown to the eavesdropper, decoding is particularly difficult. Even if $H$ is known to the eavesdropper and $G$ is invertible, the eavesdropper can construct:
\[
H G^{-1}  Y_E^n = C \lambda + H G^{-1}  N_E^n
\]

Given that $H G^{-1}$ is not unitary, such a transformation would result in correlated noise at the eavesdropper with covariance $G^{-t} H^t H G^{-1}$, which is not a scaled version of an identity matrix. As the lattice $\Lambda$'s Voronoi region is designed  to be decodable in the presence of white noise, decoding will fail at the eavesdropper.

\subsection{SVD and Inverse Water-filling}

In general, however, $H$ may not be an invertible matrix. Moreover, channel inversion at the transmission is not a good strategy in a communication system as it lowers the rates that can be supported by the medium.  A strategy that is somewhat less na\"{i}ve is presented next:

Assuming the singular value decomposition (SVD) of the channel matrix is given by:
\[
H = U D V
\]

where $U$ and $V$ are $n \times n$ unitary matrices  and $D$ is a diagonal matrix comprised of its singular values. We assume the diagonal values in $D$ are arranged in decreasing order. Given this, we can rewrite the received signal at Bob as:

\[
Y_B^n = U D V X^n + N_E^n
\]

through suitable unitary transformations, we can obtain:

\[
{\widetilde Y}_B^n = U^t Y_B^n = D {\widetilde X}^n  + N_E^n
\]

where ${\widetilde X}^n \triangleq V X^n$ . This process effectively diagonalizing the legitimate channel. If we define a minimum threshold for the channel gain $t$ below which communication is feasible, then $D$ can be subdivided into the following form:

\[
D = \left[\begin{array}{cc} { D_1} & 0 \\ 0 & {\hat D} \end{array}\right]
\] 

where ${\hat D}$ comprises of channel gains that are less than or equal to $t$. If ${\widetilde D}$ is a $k \times k$ matrix, then it is desirable that only $k$ dimensions be used for communication. A traditional wireless/multiple antenna communication system would treat these $k$ dimensions as independent parallel channels and {\em waterfill} across them \cite{Goldsmith}.   This approach is known to be optimal, but the lattice-based secrecy benefits of such a scheme are unclear. Instead, using the $k$ parallel dimensions to communicate a lattice vector makes them dependent. This dependence need not always result in a lower rate, but it can lead to the physical-layer lattice cryptosystem studied in this paper.

\noindent{\bf Encoding:} A $k$-dimensional lattice point encoded using a  lattice which permits low-complexity encoding and decoding is now used to communicate over the legitimate channel. This $k$ dimensional lattice can be obtained by truncating an existing $n$ dimensional lattice construction. This lattice point is then enhanced to $n$ dimensions by zero padding. This zero padded lattice point is denoted by  ${\tilde \lambda}$. 

\noindent{\bf Transmission:} Alice transmits $$X = C V^t {\widetilde D}^{-1}{\tilde \lambda}.$$ Here, ${\widetilde D}^{-1} $ equals
\[
{\widetilde D}^{-1} = \left[\begin{array}{cc} { D_1}^{-1} & 0 \\ 0 & {0} \end{array}\right]
\] 

and $C$ is the power normalization constant. Note that this is, in essence, truncated inverse water-filling. Such a scheme is definitely not optimal from the perspective of maximizing rate for the legitimate channel. However, it ensures that the structural properties of the lattice are maintained as it is communicated across the legitimate channel.

\noindent{\bf Reception:} Bob first constructs ${\widetilde Y}_B^n$ from $Y_B^n$ by multiplying  $U^t$, and then uses the first $k$ positions to determine the k-dimensional lattice point being communicated. As the lattice admits a low complexity decoding algorithm and is structured to handle i.i.d. Gaussian noise, this decoding is successful with high probability.

\noindent{\bf Eavesdropper:}
The eavesdropper, as before, observes a general lattice given by:
\[
Y_E = C G V^t {\widetilde D}^{-1}{\tilde \lambda} + N_E
\]

with no particular relationship between $H$ and $G$, this lattice is a general one. Even the knowledge of the encoding strategy and $H$ does not help, as any linear processing will result in colored  noise at the receiver.

Note that both the schemes presented here relied on the assumption that neither $H$ nor $G$ are a multiple of identity, i.e. constant channels. In essence, channel variation and inter-symbol interference are being used to instill an asymmetry between the legitimate receiver and eavesdropper, which is essential to the analysis. Moreover, this analysis lends itself naturally to multiple antenna channels as well. In the multiple antenna context, both static and varying channels are of interest, and this scheme is applicable to either of these settings.

\section{Construction of Lattices that Permit Efficient Encoding and Decoding}
\label{sec:block}

As mentioned earlier, the family of low density lattice codes (LDLCs) in \cite{LDLC} is one example of lattices that can be used for this lattice-based cryptosystem. Here, we summarize an alternate scheme developed in \cite{Ghiya}. Note that this serves as a summary of the unpublished work in \cite{Ghiya} only, and is not an original contribution by the authors of this paper.

This construction of the class of lattices in \cite{Ghiya} is based on the Construction-A framework  \cite{Conway}. The ensemble of Construction-A lattices is known to contain lattices that are good for source and channel coding \cite{Zamir}. Hence, this class of lattices is of particular interest to us. Unfortunately, a randomly chosen Construction-A lattice does not  afford either low-complexity encoding or decoding. Thus, a specific construction is required to enable low complexity processing.

In general, Construction-A lattices are of the form:
\[
\Lambda = p^{-1} G  {\mathbb Z}^k_p + {\mathbb Z}^n
\]

where $G$ is an $n \times k$ ``generator" matrix, ${\mathbb Z}^k_p$ is the set of all $k$-length vectors modulo $p$ and the multiplication between $G$ and ${\mathbb Z}_p^k$ is defined modulo $p$. All other operations are over the real field.

This construction is based on the understanding that a short block length is sufficient for realizing shaping gain, while longer block lengths are required to achieve coding gain. Thus, the Construction-A parity check matrix for this lattice is structured as follows:
\begin{equation*}
F = \left[\begin{array}{cccc} K & 0 & \ldots & 0 \\
A_{21} & K & \ldots & 0 \\
\vdots & \vdots & \vdots & \vdots \\
A_{l1} & A_{l2} & \ldots & K \end{array}\right]
\end{equation*}

In short, the parity-check matrix is chosen to be block lower triangular. Here $K$ is a suitable ``small" parity check matrix that provides shaping gain. For example, it could be the parity check matrix corresponding to the Leech lattice, or one corresponding to a low-density generator matrix (LDGM) code. Its sole purpose is to enable, within a dimension of a few tens, for much of the shaping gain to be captured. $A_{ij}$ are, on the other hand, chosen to be matrices such that the overall parity check matrix $F$ is ``good" from the perspective of coding gain.

Encoding for this lattice proceeds as follows: a vector $X$ is determined that solves the equation
\[
F X = \left[\begin{array}{c} 0 \\ m \end{array}\right]
\]

where $m$ is the message to be communicated. Note that this encoding procedure can result in multiple solutions for $X$ in ${\mathbb R}^n$. Thus, our interest is in determining that solution which is closest to the origin. In general, solving for a vector closest to the origin in a lattice is an NP-hard problem. However, given the structure of the lattice, this can be solved recursively. By breaking $X$ into $l$ parts:
\[
X^t = [X_1^t ~ X_2^t ~\ldots~ X_l^t]
\]

and solving consecutive equations of the form:

\[
K X_i =  \left[\begin{array}{c} 0 \\ m_i \end{array}\right]
\]

where the closest vector problem is now reduced to one over a much smaller scale than the original problem.

Decoding proceeds using conventional mechanisms in non-binary low density parity check (LDPC) codes. If $A_{ij}$ are chosen in a manner to render the code a good LDPC, a belief propagation algorithm can be used for decoding.

This concludes the summary of this particular construction. Multiple other constructions exist and are being actively researched to enable lattice-based communication, and such constructions can be used in this cyptosystem as well as non-trivial linearly transformed versions of these lattices result in general lattices over which decoding is hard.

\section{Conclusion}
\label{sec:conclude}

This paper builds a physical-layer based cryptosystem using the hardness of lattice quantization of generalized lattices. This scheme uses no key, either public or private. It uses the channel channel asymmetry between the legitimate pair and the eavesdropper in developing the scheme. Although analytically simple in design, this scheme is a natural application of lattices to both communication and secrecy. 

\section{Acknowledgment}
We thank Ankit Ghiya, Sang Hyun Lee, Krishna Narayanan and Henry Pfister for their work on constructing block lower-triangular Construction-A lattices.


\begin{thebibliography}{1}
\bibitem{Cover}
 T. Cover and J. Thomas, ``Elements of Information Theory", Springer, 1991.
 \bibitem{Wyner}
 A.D. Wyner, ``The wiretap channel", Bell System Technical Journal, 1975.
\bibitem{Shamir}
A. Shamir, ``Identity-based cryptosystems and signature schemes", Advances in Cryptology, Lecture notes in Computer Science, 1985.
\bibitem{McEliece}
R. J. McEliece, ``A public-key cryptosystem based on algebraic coding theorey", DSN Progress Report, 1978.
\bibitem{Liang}
Y. Liang, H. V. Poor and S. Shamai, ``Information theoretic security", Foundations and Trends in Communications, NOW Publishers, 2009.
\bibitem{Rhee}
M. Y. Rhee, ``Cryptography and secure communications", McGraw Hill, New York, 1993.  
\bibitem{Physical}
M. Debbah, H. El Gamal, H. V. Poor and S. Shamai (Shitz), ``Wireless physical layer security", EURASIP Journal on Wireless Communications and Networking, 2009.
\bibitem{Khot}
D. Micciancio and S. Goldwasser, ``Complexity of lattice problems", Springer, 2002.
\bibitem{Peikert}
Chris Peikert, ``Public-key cryptosystems from the worst-case shortest vector problem," in Proceedings of the 41st annual ACM symposium on Theory of computing, 2009. 
\bibitem{Zamir}
R. Zamir, ``Lattices are everywhere", Proc. ITA workshop, 2009.
\bibitem{Conway}
J. H. Conway and N. J. A. Sloane, ``Sphere packings, lattices and groups", Springer, 1988.
\bibitem{Ghiya}
A. Ghiya, S. Lee, K. Narayanan, H. Pfister and S. Vishwanath, ``Low density Construction-A lattice codes", preprint.
\bibitem{Rimoldi}
R. Urbanke, B.  Rimoldi,  ``Lattice codes can achieve capacity on the AWGN channel", IEEE Transactions on Information Theory, 1995.
\bibitem{Erez}
U. Erez and R. Zamir ,"Achieving $0.5 \log(1+SNR)$ over the additive white Gaussian noise channel with lattice encoding and decoding",  IEEE Transactions on Information Theory,  2004.
\bibitem{LDLC}
N. Sommer and M. Feder and O. Shalvi, ``Low density lattice codes", Proceedings of the IEEE International Symposium on Information Theory 2006.
\bibitem{Goldsmith}
A. Goldsmith, ``Wireless Communications", Cambridge University Press, 2005
\end{thebibliography}
\end{document}